\documentclass[twocolumn,showpacs,preprintnumbers,amsmath,amssymb]{revtex4}
\usepackage{amsthm}        
\usepackage{dcolumn}        
\usepackage{bm}
\usepackage{graphicx} 
\usepackage{subfigure}

\begin{document}

%\preprint{MPIPKS/11-2001}
%\draft

\title{Arbitrary distribution and nonlinear modal interaction in 
coupled nanomechanical resonators}

\author{J. Dorignac${}^{1,3}$, A. Gaidarzhy${}^1$, P. Mohanty${}^2$}

\affiliation{${}^1$ College  of  Engineering,  Boston University,  Boston  MA
02215\\
${}^2$ Department of Physics,  Boston University,  Boston  MA
02215\\
${}^3$ LPTA, Universit\'e Montpellier 2, 34095 Montpellier CEDEX 5}

\date{\today}
\begin{abstract}
We propose a general one-dimensional {\em continuous} formulation to analyze 
the vibrational modes of antenna-like nanomechanical resonators consisting
of two symmetric arrays of cantilevers affixed to a central nano-beam.
The cantilever arrays can have arbitrary density and length 
profile along the beam. We 
obtain the secular equation that allows for the determination
of their frequency spectrum and illustrate the results on the particular 
examples of structures with constant or alternating cantilever length profiles.
We show that our analytical results 
capture the vibration spectrum of such resonators and elucidate 
key relationships that could prove advantageous for experimental 
device performance. Furthermore, using a perturbative approach to treat 
the nonlinear and dissipative dynamics of driven structures, 
we analyze the anharmonic coupling between two specific
widely spaced modes of the coupled-element device, with direct application 
to experiments.
\end{abstract}


\maketitle

\section{Introduction} \label{Introduction}

Mechanical resonators,  especially in  the micro- and  nano-range, are
currently  used  in many  research  areas  to investigate  fundamental
physics  problems such  as  ultra-sensitive force  and mass  detection
\cite{rugar-force},    single   spin    detection   \cite{rugar-spin},
gravitational   wave   detection   \cite{bocko-onofrio}   or   quantum
measurement and computation \cite{zurek}, to  cite a few. A key aspect
of their performance is  ultra-high resonance frequency of vibration -
demonstrated up  to several gigahertz as the  nano-structures are made
from  stiffer  materials  such  as  UNC  diamond  or  silicon  carbide
\cite{huang,imboden}, and designed with submicron critical dimensions.
The coupling of arrays of high frequency resonators can yield enhanced
response  characteristics in  hybrid  resonator designs,  and as  such
holds primary  interest in  a plethora of  technological applications.
Detailed  analytical   and  numerical  modeling   of  coupled  element
resonators is  critical to the  understanding and utilization  of such
devices.

In a previous publication \cite{dorignac_JAP1} we presented both discrete
and  continuum models  for a  class of  coupled-element nanomechanical
antenna-like  structures   that  have  been   experimentally  realized
\cite{gaidarzhy-prl,gaidarzhy-apl,imboden}.   The continuum  model was
shown to capture  the linear modal response spectrum  of the device in
close agreement with finite  element simulations.  As a sequel to the previous work, here we generalize
the analysis  to coupled-element structures  with arbitrary cantilever
lengths and distributions along the central beam.  By judicious choice
of  cantilever  array  distribution,  the  response  spectrum  of  the
structure can be  engineered for specific performance characteristics.
As an illustration, we treat the particular cases of constant and 
alternating cantilever length arrays. Motivated by  experimental  
observations  of  modal coupling phenomena, we further expand the 
analysis to include weak nonlinear and dissipative contributions
to the dynamics of a driven cantilever-beam and derive the 
amplitude-frequency response that describes the anharmonic coupling  
taking place between two of its specific widely spaced modes.

\section{Arbritrary Cantilever Distribution} \label{sec:arbitrary}

\begin{figure}
\includegraphics[width=0.85\linewidth,height=1.1\linewidth]{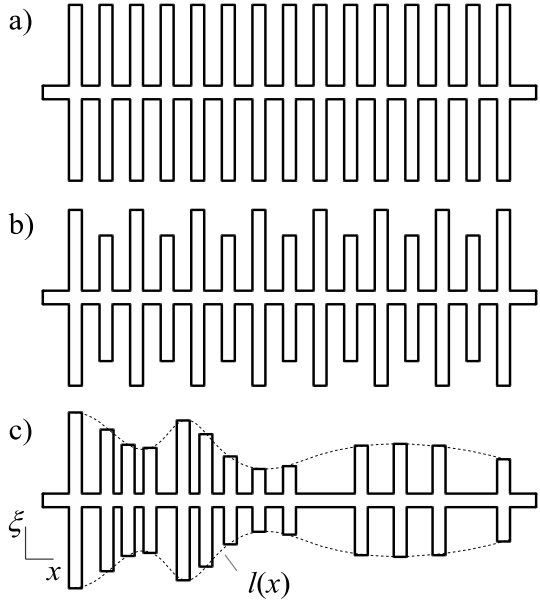}\caption{\label{FigAnt}  Schematic  (2D)  top  views  of mechanical
antenna  structures   with  symmetric  arrays   of  cantilevers  (along
$\xi$-axis). a) Constant lengths. b) Alternating lengths. c) Length profile, $l(x)$, and density along the central beam ($x$-axis) 
are arbitrary.  Deflections occur perpendicular to the $(x,\xi)$-plane.}
\end{figure}

We consider a doubly-clamped elastic beam with two symmetric arrays of
cantilevers with lengths $l(x)$ attached on both  sides, see Fig. \ref{FigAnt}.  We denote
the beam  out-of-plane deflection  by $y(x,t)$, $x  \in [0,L]$  and by
$\eta(x,\xi,t)$,  the  deflection  with  respect to  $y(x,t)$  of  the
cantilevers  located  at  point  $x$  on  the  beam.  Restricting  our
investigation to  modes symmetric  with respect to  the beam,  we have
$\xi \in [0,l(x)]$.

With subscripts  $b$ and  $c$ referring to  quantities related  to the
beam  and  the cantilevers,  respectively,  we  denote the  rigidities
\footnote{The rigidity of an elastic element is given by ${\cal E}=EI$
where $E$  is Young's  modulus and  $I$ the moment  of inertia  of the
element.  See for example  \cite{Gere}.}  by  ${\cal E}_b$  and ${\cal
E}_c$,  the masses  per unit  length by  $\mu_b$ and  $\mu_c$  and the
widths by $w_b$ and $w_c$. If we assume that all elements have the same
thickness,  ${\cal    E}_b/{\cal   E}_c=\mu_b/\mu_c=w_b/w_c$
\cite{dorignac_JAP1}.  The cantilever  density is defined as $\rho(x)=
dN(x)/dx$, where $N(x)$  is the total number of  cantilevers between 0
and $x$.   Assuming that shear  deformation and rotary inertia  of the
structure  are negligible,  we work  within the  Euler-Bernouilli beam
approximation.    Although  not  essential   to  our   approach,  this
assumption greatly simplifies the interpretation of the results.  More
sophisticated   beam   theories  can   for   instance   be  found   in
Ref.  \cite{Han99}  and   references  therein.   The  Euler-Bernouilli
Lagrangian of  the structure depicted in Fig.   \ref{FigAnt}c is given by

\begin{multline} \label{Lagrangian}
{\cal L}[y,\eta](t) 
= \int\limits_0^L dx \left[\frac{\mu_b}{2}\left(\frac{\partial y}
{\partial t}\right)^2-\frac{{\cal E}_b}{2}
\left(\frac{\partial^2 y}{\partial x^2}\right)^2 \right] + \\
\int\limits_0^L \!dx \, \rho(x) \int\limits_0^{l(x)} d\xi 
\left[\frac{\mu_c}{2}\left(\frac{\partial \eta}{\partial t}+
\frac{\partial y}{\partial t}\right)^2 - 
\frac{{\cal E}_c}{2}\left(\frac{\partial^2 \eta}{\partial \xi^2}\right)^2 
\right].
\end{multline}
Cantilever boundary conditions apply to the deflection $\eta(x,\xi,t)$ and 
read 
\begin{equation} \label{BCC}
\left.\eta\right|_{\xi=0}=
\left.\frac{\partial \eta}{\partial \xi}\right|_{\xi=0}=
\left.\frac{\partial^2 \eta}{\partial \xi^2}\right|_{\xi=l(x)}=
\left.\frac{\partial^3 \eta}{\partial \xi^3}\right|_{\xi=l(x)}=0.
\end{equation}

Boundary conditions for the beam can be arbitrarily specified. 
For a clamped-clamped beam, for instance, they would be 
$y(0,t)=y'(0,t)=0$ and $y(L,t)=y'(L,t)=0$. 
From the least action principle, 
$\delta {\cal S}[y,\eta] = 0$, where 
${\cal S}[y,\eta] = \int \! {\cal L}[y,\eta](t) \, dt$, we obtain 
the following equations of motion
\begin{eqnarray}
&& {\cal E}_b \frac{\partial^4 y(x,t)}{\partial x^4} + \mu_b 
\frac{\partial^2 y(x,t)}{\partial t^2} = - {\cal E}_c\rho(x)
\left. \frac{\partial^3 \eta(x,\xi,t)}{\partial \xi^3} 
\right|_{\xi=0}  \label{Eqmoty}\\
&& {\cal E}_c \frac{\partial^4 \eta(x,\xi,t)}{\partial \xi^4} + \mu_c 
\frac{\partial^2 \eta(x,\xi,t)}{\partial t^2} = -\mu_c 
\frac{\partial^2 y(x,t)}{\partial t^2}, \label{Eqmoteta} 
\end{eqnarray}
The right hand side of Eq. \eqref{Eqmoty} is the shear force exerted
by the cantilevers along the beam. Eq. \eqref{Eqmoteta} combines 
the equations of motion of the cantilevers in a single one via 
$\eta(x,\xi,t)$. Its right hand side is the inertial 
driving force exerted by the central beam on the cantilevers. The energy 
of system \eqref{Eqmoty}-\eqref{Eqmoteta}, whose expression 
is obtained from Eq. \eqref{Lagrangian} by
changing the sign of the elastic energy terms, is a constant of the motion.

We now look for mode solutions of the form
\begin{equation} \label{modesol}
y(x,t)=Y(x)\cos \omega t\ ;\ \eta(x,\xi,t) = H(x,\xi)\cos \omega t.
\end{equation}
Defining the useful quantities $\alpha$ and $\gamma$ by
\begin{equation} \label{alphagamma}
\alpha^4 = \frac{\mu_b}{{\cal E}_b}\omega^2\ ; \ \gamma = \alpha l(x),
\end{equation}
we can solve Eq. \eqref{Eqmoteta} together with boundary conditions 
\eqref{BCC} to obtain
\begin{multline} \label{Hsol}
H(x,\xi) = \left[A_1^+(\gamma) \cos(\alpha \xi) + 
A_2^+(\gamma)\sin(\alpha \xi) + \right. \\
\left. A_1^-(\gamma)\cosh(\alpha \xi) + 
A_2^-(\gamma) \sinh(\alpha \xi) - 1\right] Y(x) 
\end{multline} 
where 
\begin{eqnarray}
&& A_1^{\pm}(\gamma) = \frac{1+\cos \gamma \cosh \gamma \mp 
\sin \gamma \sinh \gamma}{2(1+\cos \gamma \cosh \gamma)}, \label{a1a3}\\
&& A_2^{\pm}(\gamma) = \pm 
\frac{\cos \gamma \sinh \gamma + 
\sin \gamma \cosh \gamma}{2(1+\cos \gamma \cosh \gamma)}. \label{a2a4} 
\end{eqnarray}
Using Eq. \eqref{modesol} again and Eqs. \eqref{Hsol}-\eqref{a2a4}, we can 
evaluate the right hand side of the equation of motion for the beam,
$\partial^3 H/\partial \xi^3|_{\xi=0}=-2\alpha^3A_2^+(\gamma)Y(x)$ 
and Eq. \eqref{Eqmoty} can finally be cast into the form (the superscript
$(iv)$ denotes a fourth derivative)
\begin{equation} \label{Eqmotymode}
Y^{(iv)}(x)-\left[V(\alpha;x)+\alpha^4\right]Y(x)=0,
\end{equation}
where the ``potential'' $V(\alpha;x)$ is given by
\begin{equation} \label{Valphax}
V(\alpha;x)= \frac{w_c}{w_b}\rho(x)\alpha^3
\frac{\cos \gamma \sinh \gamma + 
\sin \gamma \cosh \gamma}{1+\cos \gamma \cosh \gamma}.
\end{equation}
The frequency spectrum of the antenna is obtained by solving
Eq. \eqref{Eqmotymode} with four boundary conditions for the beam
deflection $Y(x)$ which results in a secular equation for the parameter 
$\alpha$ that is, for the frequency $\omega$ (see Eq. \eqref{alphagamma}).
For example, if cantilevers and beam alike are assumed to be 1D-like, 
the density reads $\rho(x)=2\sum_j \delta(x-x_j)$, 
where the $x_j$'s are the locations of the cantilevers 
along the beam.  
The peculiarity of Eq. \eqref{Eqmotymode} is to have an $x$-dependent
potential whenever the density or the length profile vary along the beam.
Unless an explicit solution is known, it can be solved by 
projection onto an orthonormal 
set of modes satisfying appropriate boundary conditions for the beam.
Eq. \eqref{Eqmotymode} then becomes an homogenous matrix equation 
whose determinant provides the secular equation that
yields the frequency spectrum of the structure.

\section{Constant and alternating cantilever length arrays} \label{sec:bidim}

{\em Constant cantilever length -}
We now turn to the special case depicted in Fig. 
\ref{FigAnt}a where two dense arrays of $N$ regularly 
spaced cantilevers with constant length $l$ are located symmetrically 
with respect to the central beam. 
As previously stated, in a one-dimensional setting, their exact density 
is twice the sum of $N$ equispaced delta functions. 
When $N$ is large enough, it is physically relevant to 
replace the density by its average value over the beam, $\bar{\rho}=2N/L$, 
as long as the modal shapes of the beam under investigation wash out 
the discreteness of the cantilever distribution.
This requires for instance that the number of nodes of 
the said modal shapes, $n$,
be small compared to the number of cantilevers on one side, $N$. Hence 
the condition of validity of our approach, $n \ll N$. 

The physical
meaning of replacing the exact density by its average is to spread 
the local shear forces exerted by the cantilevers on the beam around their 
support in order to form a continuous force density. 
This results in a {\em cantilever continuum} represented by the continuous
deflection $\eta(x,\xi,t)$. Notice that, although bi-dimensional, 
the cantilever continuum is distinct from an elastic plate because
adjacent cantilevers are not elastically connected. 
As $\rho(x)$ is now replaced by $\bar{\rho}$ and $\gamma=\alpha l$ is a 
constant, the ``potential'' $V(\alpha;x)$ does not depend on $x$ anymore
and the solution to Eq. \eqref{Eqmotymode} is straightforward. 
Its modes are given by 
\begin{equation} \label{Ymoderholconst}
Y_n(x)= A \varphi_n(x/L),
\end{equation}         
where $A$ is an overall amplitude and $\{\varphi_n(u)\}$, 
an orthonormalized set of modes satisfying 
\begin{equation} \label{varphin}
\varphi^{(iv)}_n(u)=\beta_n^4\varphi_n(u),\ \int\limits_0^1\! du\, 
\varphi_n(u)\varphi_m(u) = \delta_{m,n}.
\end{equation} 
%\begin{figure}
%\includegraphics[width=0.85\linewidth,height=0.75\linewidth]
%{Spectrum4pa05N20lL005bis.eps}
%\caption{\label{FigSpec} Frequency spectrum of an antenna with
%$N=20$ cantilevers on one side, length ratio $l/L=0.05$ and width
%ratio $w_c/w_b=0.5$. Frequencies $\omega_{n,k}$ are
%normalized to the fundamental's, $\omega_{1,1}$. Physically relevant 
%levels (such that $n<N$) are represented as solid lines and others 
%as dashed lines.}
%\end{figure}
\begin{figure}
\includegraphics[width=0.85\linewidth,height=0.75\linewidth]
{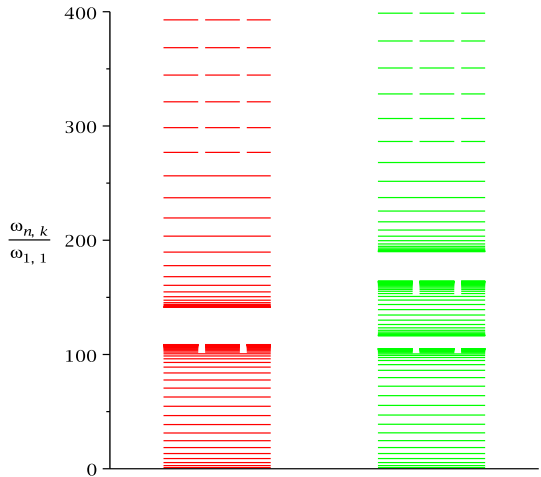}
\caption{\label{FigSpec} Frequency spectra 
of a constant cantilever length structure (left) and of 
an alternating cantilever length structure (right). 
Frequencies $\omega_{n,k}$ are normalized to the fundamental's, 
$\omega_{1,1}$. Physically relevant 
levels (such that $n<N$) are represented as solid lines and others 
as dashed lines. Other parameters are given in the text.}
\end{figure}
The $\beta_n$'s are the positive roots of the secular equation 
obtained from the 
boundary conditions for the central beam. For instance, for a 
clamped-clamped $(+)$ or a clamped-free $(-)$ beam,
\begin{equation} \label{secbeta}
\cos \beta_n \cosh \beta_n = \pm 1.
\end{equation}  
The secular equation for $\alpha$, conveniently expressed here in terms 
of $\gamma=\alpha l$, 
is obtained from Eqs. \eqref{Eqmotymode} and \eqref{Ymoderholconst} as
\begin{equation} \label{secgamma}
\frac{2Nm_c}{m_b}\gamma^3
\frac{\cos \gamma \sinh \gamma + 
\sin \gamma \cosh \gamma}{1+\cos \gamma \cosh \gamma}+\gamma^4 = 
\left(\frac{l\beta_n}{L}\right)^4,
\end{equation}
where $m_b=\mu_b L$ and $m_c=\mu_c l$ denote the mass of the beam and of 
a cantilever, respectively. For each particular value of $\beta_n$, 
Eq. \eqref{secgamma} has an 
infinite number of roots $\gamma_{n,k}$, $k\geq 1$, from which the frequency
is evaluated as
\begin{equation} \label{omegank}
\omega_{n,k} = \sqrt{\frac{{\cal E}_c}{\mu_c}}
\left(\frac{\gamma_{n,k}}{l}\right)^2.
\end{equation}
The corresponding modes are 
determined from Eqs. \eqref{Ymoderholconst} and 
\eqref{Hsol} where $\gamma$ is replaced by its value $\gamma_{n,k}$. 
As we see from Eq. \eqref{Ymoderholconst}, the modal shape 
of the beam is the same for the antenna as it would be for 
a beam {\em without cantilevers attached}, a result also confirmed 
by finite element analysis \cite{dorignac_JAP1}. 
This surprising fact originates in
the special form of the force density exerted by the cantilevers 
on the beam that is proportional to the beam deflection $Y_n(x)$, 
thereby producing no change in the modal shape even though the 
frequency spectrum is modified by the first term of Eq. 
\eqref{secgamma} for $N>0$. 

Eq. \eqref{secgamma} has three relevant parameters:  
the length ratio, $\lambda=l/L$, the product of the cantilever number 
by the width ratio, $\nu=2Nw_c/w_b$ and $\beta_n$. 
It can be shown that, for all $n\geq 1$ and $k\geq 2$, its solutions satisfy  
$\gamma_{\infty,k-1}< \gamma_{n,k}<\gamma_{\infty,k}$ where 
$\gamma_{\infty,k}$ satisfies the secular equation of a simple 
cantilever,
\begin{equation} \label{gammainfk} 
\cos \gamma_{\infty,k} \cosh \gamma_{\infty,k}+1=0.
\end{equation} 
Notice that although the values $\gamma_{\infty,k}$ are {\em parameter 
independent}, the corresponding frequencies, $\omega_{\infty,k}$, 
derived from \eqref{omegank} still depend on $l$ and other physical 
quantities. The frequencies satisfy
$\omega_{\infty,k-1}< \omega_{n,k}<\omega_{\infty,k}$, which 
explains the generic ``band'' structure of the spectrum illustrated 
in Fig. \ref{FigSpec} (right) on the specific example of an antenna 
clamped at both ends with $N=20$ cantilevers, $w_c=w_b$ and $l/L=0.05$. 

To gain some insight in the fine structure of the bands, we observe 
that Eq. \eqref{secgamma} can be solved approximately for large 
$n$ and for small or large $\lambda$. These limits all require its 
first term to become large for finite values of
$\gamma$, which leads to $\gamma_{n,\tilde{k}} = 
\gamma_{\infty,k} +\Delta_{n,k}$ with  
\begin{equation} \label{Deltank}
\Delta_{n,k} \simeq
\frac{\lambda\nu}{\gamma_{\infty,k}}\frac{t_{\infty,k}+th_{\infty,k}}
{t_{\infty,k}-th_{\infty,k}}
\left[1-\left(\frac{\lambda\beta_n}{\gamma_{\infty,k}}\right)^4\right]^{-1}.
\end{equation} 
In this expression, $t_{\infty,k}\equiv \tan \gamma_{\infty,k}$, 
$th_{\infty,k}\equiv \tanh \gamma_{\infty,k}$; $\tilde{k}=k$ 
if $\Delta_{n,k}<0$ and $\tilde{k}=k+1$ if $\Delta_{n,k}>0$. Eq. 
\eqref{Deltank} is valid for $|\Delta_{n,k}|\ll 1$, $k\geq 1$ 
\footnote{For a more accurate version of this result, see Ref. 
\cite{dorignac_JAP1}}. 

We can show that regardless of the boundary conditions, $\Delta_{n,k}$ 
vanishes as $n$ becomes large and the frequencies $\omega_{n,k}$ 
cluster at $\omega_{\infty,k}$, the {\em upper edge} of band $k$.
Provided $\lambda$ is small enough, Eq. \eqref{Deltank} indicates 
that an accumulation of frequencies also occurs at the {\em lower edge} 
of band $k+1$, a phenomenon observed in Fig. \ref{FigSpec}. In this limit, 
the frequency band-gap, 
$\Delta\omega_k \equiv \omega_{1,k+1}-\omega_{\infty,k}$ is given by
$\Delta\omega_k\simeq 2\omega_{\infty,k}\Delta_{1,k}/\gamma_{\infty,k}$. 
It is seen to become constant as $k$ increases.
\begin{figure}[t]
\includegraphics[width=1.\linewidth,height=0.75\linewidth]
{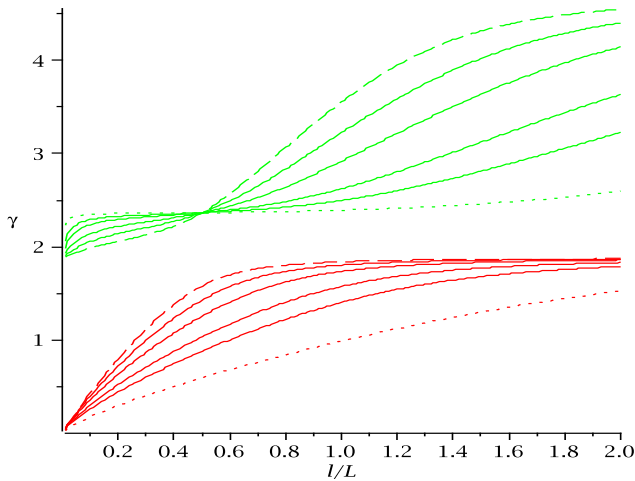}
\caption{\label{Figgamma11n12vslLN} Fundamental, $\gamma_{1,1}$ (red), and 
first collective, $\gamma_{1,2}$ (green), levels versus the length ratio
$\lambda = l/L$ for $N=5,10,20,50,100$ and $500$. 
For clarity, $N=5$ and $N=500$ are displayed as dashed and dotted 
lines, respectively.}
\end{figure} 
Clearly, from Eq. \eqref{Deltank}, if $n$ is fixed and $k\geq 2$, 
$\gamma_{n,k}$ tends to $\gamma_{\infty,k-1}$ ($\gamma_{\infty,k}$) 
when the length ratio $\lambda$ tends to zero (infinity).
This indicates that, for band 2 
and higher, the cantilever dynamics (see Eq. \eqref{gammainfk}) 
dominate the motion of the structure in these two limits. This is not true
however for the first (fundamental) band whose lower edge tends to zero
as $\lambda \rightarrow 0$ and whose dynamics is 
governed by the beam's in this limit, a mass loading effect.

An interesting example of level motion is given in 
Fig. \ref{Figgamma11n12vslLN} where 
the fundamental mode, $\gamma_{1,1}$, and the 
first collective mode, $\gamma_{1,2}$, are plotted against the length 
ratio $\lambda$ for various cantilever numbers, $N$.
The central beam is again clamped at both ends and the width
ratio is $w_c/w_b=0.5$ ($\nu=N$). 
For all $N$, $\gamma_{1,1}$ interpolates between
zero and $\gamma_{\infty,1}$ as $\lambda$ increases and we note that,
for a given length ratio, the fundamental frequency {\em always 
shifts downwards} when cantilevers are added to the central beam. 
In the small $\lambda$ regime, this is the manifestation 
of a {\em mass loading} effect while for large $\lambda$, this 
is predicted by Eq. \eqref{Deltank}. 
Similarly, $\gamma_{1,2}$ interpolates beween 
$\gamma_{\infty,1}$ and $\gamma_{\infty,2}$ but there is a 
special ratio, $\lambda^*_{1,2}$, for which it
becomes $N$-{\em independent} (curve intersection
in Fig. \ref{Figgamma11n12vslLN}). For 
$\lambda < \lambda^*_{1,2}$, $\gamma_{1,2}$ increases with $N$ and
decreases otherwise, as predicted by \eqref{Deltank} 
for small and large $\lambda$, respectively.  

Such a ratio exists in fact for any level $\gamma_{n,k}$, $k\geq 2$. 
It is given by 
$\lambda^*_{n,k}=\beta^{cc}_{2k-3}/(2\beta_{n})$ where $\beta^{cc}_m$ 
is solution to the clamped-clamped beam equation \eqref{secbeta}
while $\beta_n$ is solution to the actual beam secular equation. 
For that ratio, 
$\gamma^*_{n,k}=(l \beta_n)/L$. Using 
${\cal E}_b/\mu_b={\cal E}_c/\mu_c$ and Eq. \eqref{omegank}, 
the corresponding frequency is thus 
$\omega^*_{n,k}=\sqrt{{\cal E}_b/\mu_b}(\beta_n/L)^2$, which is precisely 
the $n$th frequency of a beam {\em without cantilevers attached!} For the 
clamped-clamped antenna of Fig. \ref{Figgamma11n12vslLN}, 
$\lambda^*_{1,2}=1/2$ and $\omega^*_{1,2}=
\sqrt{{\cal E}_b/\mu_b}(\beta_1/L)^2$. 
Thus, if the cantilever length is half the beam's,
the 1st collective frequency of the structure does not depend on the number
of cantilevers attached and is equal to the fundamental frequency of the
central beam alone. Although this conclusion is drawn from 
our continuum approach, it has been verified for the dynamics 
of the exact 3D structure solved by the finite-element method. 

% beginning of previously commented text

% The existence of a length ratio 
% for which a level $\gamma$ is $N$-independent is not specific to 
% $\gamma_{1,2}$. It occurs for all level $\gamma_{n,k}$ with 
% $n \geq 1$ and $k\geq 2$, and is determined by the two conditions 
% $\tan \gamma_{n,k} + \tanh \gamma_{n,k}=0$ and 
% $\gamma_{n,k} = \lambda \beta_n$. It is possible to show that the exact
% solution to the first equation is $\gamma_{n,k}=\beta^{cc}_{2k-3}/2$ where
% $\beta^{cc}_m$ is solution to the clamped-clamped beam equation given in
% Eq. \eqref{secbeta}. For it 
% to be compatible with the second condition one must have 
% $\lambda=\lambda^*_{n,k}=\beta_{2k-3}/(2\beta_{n})$.  
 
We also note that, 
all other parameters being fixed, the density of states (number of states 
per unit frequency) increases with $N$ and that, for $k\geq 2$, the lower 
band edges generically shift upwards. The opposite occurs however
for the first band (fundamental group) whose lower edge shifts downwards
as $N$ increases. The role of the cantilevers is indeed different for the 
fundamental group ($k=1$), where they essentially produce a mass loading 
of the central beam, than for higher bands where their dynamics take over and
shift the antenna frequencies upwards. 

{\em Alternating cantilever length arrays -}
We briefly comment here on the case of arrays of cantilevers with alternating 
lengths, $l_1$ and $l_2$, $l_2 \leq l_1$ (Fig. \ref{FigAnt}b). 
Using the continuum approach, we obtain the corresponding 
secular equation   
\begin{equation} \label{secgamma2l}
\gamma^4 +\frac{w_{1}l_1\bar{\rho}_1}{w_b} 2 A_2^+(\gamma)+
\frac{w_{2}l_1\bar{\rho}_2}{w_b} 2 A_2^+(\epsilon \gamma)= 
\left(\frac{l_1\beta_n}{L}\right)^4
\end{equation} 
where $\gamma = \alpha l_1$, $\epsilon=l_2/l_1$ 
and where $\alpha$ and $A_2^+$ are defined by Eqs. 
\eqref{alphagamma} and \eqref{a2a4}, respectively. $w_{1}$ and $w_{2}$
refer to the widths of the cantilevers and $\bar{\rho}_1=2N_1/L$,
$\bar{\rho}_2=2N_2/L$ to their average density, $N_i$ being the number of 
cantilevers with length $l_i$ on one side of the structure. 
Notice that for $l_1=l_2$, Eq. \eqref{secgamma2l} reduces to 
Eq. \eqref{secgamma} as it should. 
The spectrum calculated from Eq. \eqref{secgamma2l} 
is displayed in Fig. \ref{FigSpec} (left). For comparison with 
its constant length counterpart, we have assumed $w_{1}=w_{2}=w_b$, 
$\bar{\rho}_1=\bar{\rho}_2=N/L$ with $N=20$ 
and  $l_1/L=0.05$ while $\epsilon=0.8$.  
The spectrum of the alternating structure shows additional bands. They are 
due to the term $A_2^+(\epsilon \gamma)$ that becomes
infinite when $\gamma = \gamma_{\infty,k}/\epsilon$, bands ensuing as 
previously discussed. This example shows how the frequency spectrum of a more
complex antenna structure can be easily understood and provides a way 
of engineering frequency spectra with specific features useful in 
experiments.

%end of previously commented text

\section{Nonlinear system driven by two frequencies} \label{sec:NLN2F}

\subsection{Motivation}

%It is relevant to evaluate the Lagrangian ${\cal L}_{n,k}(t)$ 
%of the system and in 
%particular its time average that we will use in the next section to 
%evaluate the effects of small nonlinearities and damping on the structure.
%We find
%\begin{equation} \label{Lagrangenk}
%{\cal L}_{n,k}(t) = -\frac{M_{n,k} A^2 \omega^2_{n,k}}{2}\cos (2\omega_{n,k}t)
%\ \ \Rightarrow \ \ \langle {\cal L}_{n,k}(t) \rangle = 0,
%\end{equation}
%where $\langle \dots \rangle$ denotes the time average. The modal Lagrangian 
%of the linear continuum model is then similar to a harmonic oscillator's and
%its time average is zero. This result will be used in the perturbative 
%treatment of a weakly nonlinear and dissipative antenna in section 
%\ref{NLN2F}. 

It has been observed experimentally \cite{gaidarzhy_thesis} that, 
if in addition to being driven at the frequency of the fundamental mode 
the antenna is 
also driven at the frequency of the collective mode, the frequency 
peak of the fundamental mode experiences a slight shift. This frequency
shift is the signature of a modal coupling that occurs because of the
presence of nonlinearity and dissipation in the system \cite{Nayfeh04}. 
To explain the interaction of these two widely spaced modes, 
we investigate the effect of a two-frequency driving on the response 
of a weakly nonlinear and dissipative antenna structure. 
We supplement the equations of motion of our continuum model, Eqs. 
\eqref{Eqmoty} and \eqref{Eqmoteta}, with 
nonlinear terms that take into account the possible 
material and geometrical nonlinearities of the structure and with damping 
terms proportional to the transverse velocity of the elements. 
These terms, small compared to the amplitude of vibration of 
the antenna, are treated within the Lagrangian approach described 
in \cite{Nayfeh04} to derive the frequency-amplitude relation of the model.

\subsection{Lagrangian approach}

It is convenient at this stage to define the mode $(n,k)$ of 
the constant cantilever length structure by the 2-column vector
$\psi_{n,k} = (y_{n}(x,t),\eta_{n,k}(x,\xi,t))^T$. It involves
two parameters, $\beta_n$ and $\gamma_{n,k}$, respective solutions 
to the secular equations Eqs. \eqref{secbeta} and \eqref{secgamma}.
The corresponding frequency $\omega_{n,k}$ is determined by Eq. \eqref{omegank}
while its modal shape is given by Eqs. \eqref{Hsol} and \eqref{Ymoderholconst}.
In particular, the fundamental mode is given by the
parameters $(\beta_1,\gamma_{1,1})$ and the collective mode we are 
interested in by the parameters $(\beta_1,\gamma_{1,2})$. For clarity
in the notations,
we rename these two modes $\psi_1 = (y_1(x,t),\eta_{11}(x,\xi,t))^T$
and $\psi_2 = (y_2(x,t),\eta_{21}(x,\xi,t))^T$ and denote
their respective frequencies by $\omega_1$ and $\omega_2$ where
\begin{equation} \label{Omega1and2}
\omega_k = 
\sqrt{\frac{{\cal E}_c}{\mu_c}}\left(\frac{\gamma_{1,k}}{l}\right)^2, 
\ \ k \in \{1,2\}.
\end{equation}
Experimental data (to be published elsewhere) shows that 
$\psi_1$ and $\psi_2$ are coupled in the sense
that the amplitude-frequency curve (resonance) of the fundamental mode
is modified when the higher mode is driven. This coupling is attributed 
to the presence of nonlinearities in the resonator structure. 
To explain this phenomenon,
we treat the nonlinearities and the damping affecting the system as a 
perturbation of the fundamental and collective modes, $\psi_1$ and 
$\psi_2$. These
perturbative terms are responsible for a modulation of the linear modes
that we evaluate by a multiple scale method. Hereafter, we closely follow 
the Lagrangian approach of Ref. \cite{Nayfeh04} because it offers 
a particularly suitable framework to derive the modulation equations. 
    
 When the driving amplitude 
(or power) is small enough, the solution $\psi = (y(x,t),\eta(x,\xi,t))^T$
to the nonlinear equations of 
motion is, to a good approximation, given by a superposition of $\psi_1$ 
and $\psi_2$ with slowly modulated amplitudes. 
Following the multiple scale approach, we write $\psi$ as 
\begin{multline} \label{yetamultscal}
\begin{pmatrix} 
y(x,t) \\
\eta(x,\xi,t)
\end{pmatrix}
= \varepsilon\left\{ A_1(T_2)\varphi_1(u)\, e^{i\omega_1 T_0}\begin{pmatrix} 
1 \\
\chi_1(v)
\end{pmatrix} + \right. \\ \left. 
+ A_2(T_2)\varphi_1(u)\, e^{i\omega_2 T_0}\begin{pmatrix} 
1 \\
\chi_2(v)
\end{pmatrix} + c.c. \right\}
\end{multline}    
In this expression, $c.c.$ denotes the complex conjugate, 
$\varepsilon$ is a small bookkeeping parameter, $u=x/L$, $v=\xi/l$,
$\chi_k(v)=H_{1,k}(x,\xi)/Y_1(x)$ and $\varphi_1(u)$ is defined in Eq. 
\eqref{varphin}. According to the multiple scale method,
we have introduced two time scales, $T_0=t$ and $T_2=\varepsilon^2 t$. The 
complex slowly varying modulation amplitudes are denoted by $A_1(T_2)$ and
$A_2(T_2)$.

The Lagrangian of our system is given by
\begin{multline} \label{LagrangianNL}
{\cal L} = \int\limits_0^L dx 
\left[\frac{\mu_b}{2}\left(\frac{\partial y}{\partial t}\right)^2 -
\frac{{\cal E}_b}{2}\left(\frac{\partial^2 y}{\partial x^2}\right)^2  
\right] + \\ +
\frac{2N}{L} \int\limits_0^L dx \int\limits_0^l d\xi 
\left[\frac{\mu_c}{2}\left(\frac{\partial \eta}{\partial t}+
\frac{\partial y}{\partial t}\right)^2 - 
\frac{{\cal E}_c}{2}\left(\frac{\partial^2 \eta}{\partial \xi^2}\right)^2 
\right] + \\ 
+ ({\rm NLT}) + \left(F_1 \cos(\Omega t) + 
F_2 \cos(\omega t) \right) \int\limits_0^L y dx,
\end{multline} 
where $({\rm NLT})$ stands for ``Nonlinear Terms''. 
To express the fact that the driving frequencies, $\Omega$ and $\omega$, 
are close to the linear frequencies of the fundamental and excited modes, 
we write them as
\begin{equation} \label{FreqDriv}
\Omega = \omega_1 + \varepsilon^2 \sigma_1\ \ ;\ \ \omega = \omega_2 + 
\varepsilon^2 \sigma_2.
\end{equation}   
To describe the nonlinear response of the cantilevers and the central beam, 
neglecting the effects of rotary inertia and shear deformations, we
add the following nonlinear terms to the Lagrangian \cite{Nayfeh04}
\begin{multline} \label{NLterms}
\hskip-2ex {\rm NLT} \!=\! \int\limits_0^L \! dx \left[ \! \frac{\mu_b}{8}\! 
\left[\frac{\partial}{\partial t}\! 
\int\limits_0^x \!
\left(\frac{\partial y}{\partial x'}\right)^2 \!\!dx' \right]^2
\!\!\!-\! \frac{{\cal E}_b}{2} 
\left(\frac{\partial y}{\partial x}\,
\frac{\partial^2 y}{\partial x^2}\right)^2 \!\right]+
\\
\frac{2N}{L}\!\int\limits_0^L \!\! dx\!\!
\int\limits_0^l \! d\xi \left[ \!
\frac{\mu_c}{8} \!\! \left[\frac{\partial}{\partial t} 
\int\limits_0^\xi \!
\left(\frac{\partial \eta}{\partial \xi'}\right)^2 \!\!d\xi' \right]^2
\!\!\!\!-\! \frac{{\cal E}_c}{2} \left(\frac{\partial \eta}{\partial \xi}\,
\frac{\partial^2 \eta}{\partial \xi^2}\right)^2 \!\right].
\end{multline} 
Finally, we take into account the damping effects of the viscous 
forces acting on the antenna through the virtual
work
\begin{multline} \label{damping}
\delta W = - \int\limits_0^L \! dx \, 
C_y \frac{\partial y}{\partial t} \delta y 
\\
- \frac{2N}{L}\int\limits_0^L \! dx
\int\limits_0^l \! d\xi \, C_{\eta} \left(\frac{\partial y}{\partial t} + 
\frac{\partial \eta}{\partial t} \right) \delta \eta,
\end{multline}   
where $C_y$ and $C_{\eta}$ are the viscosities of the beam and the cantilevers,
respectively. 

To account for the fact that damping effects and driving forces are 
of the same order of magnitude as the nonlinear effects, we scale
the viscosities as $C_{y}=\varepsilon^2 c_{y}$ and 
$C_{\eta}=\varepsilon^2 c_{\eta}$ and the forces as $F_j = \varepsilon^3 f_j$,
$j=1,2$.
We now proceed as explained in Ref. \cite{Nayfeh04} to derive a time-averaged 
Lagrangian from Eqs. \eqref{yetamultscal}, \eqref{LagrangianNL} and 
\eqref{NLterms}. We substitute \eqref{yetamultscal} into the Lagrangian 
\eqref{LagrangianNL} and also into the virtual work \eqref{damping}, perform
the spatial integrations and keep the slowly varying terms only
- i.e. those that are either constant or function of $T_2$ only. This yields:
\begin{multline} \label{LagA}
\frac{\langle {\cal L} \rangle}{\varepsilon^4} = 
\sum_{j=1}^2 \left\{i M_j\omega_j \left(A_j \bar{A_j'}- A_j'\bar{A_j}\right)
+ C_{jj}|A_j|^4 + \right. \\ \left. + 
{\cal F}_j \left(\bar{A_j}e^{i\sigma_j T_2}+cc\right)\right\} 
+ 2 C_{12}|A_1|^2|A_2|^2 
\end{multline}
\begin{equation} \label{dWQ}
\frac{\langle \delta W \rangle}{\varepsilon^4} = \sum_{j=1}^2
Q_j \delta A_j + cc,\ \ \ {\rm with}\ \ \ Q_j = 2 i \omega_j \mu_j \bar{A_j},
\end{equation}
where
\begin{eqnarray} \label{Param}
M_j &=& m_b+2Nm_t L_{jj}\ \ ;\ \ {\cal F}_j = \frac{L}{2}f_j \Gamma_4 \nonumber \\
\mu_j &=& \frac{1}{2}( L c_y + 2Nl c_{\eta} \Lambda_{jj} ) \nonumber \\
C_{jj} &=& \frac{\Gamma_1 m_b \omega_j^2}{L^2} - \
\frac{3\Gamma_2{\cal E}_b}{L^5} +\frac{2N\Gamma_3}{l^2}
\left[m_t\omega_j^2I_{jj}-\frac{3{\cal E}_t}{l^3}K_{jjjj}\right] 
\nonumber \\
C_{12} &=& \frac{\Gamma_1 m_b (\omega_1^2+\omega_2^2)}{L^2}\!-\! 
\frac{6\Gamma_2{\cal E}_b}{L^5} \!+\! \frac{2N\Gamma_3}{l^2}
\Big[m_t(\omega_1^2+\omega_2^2)I_{12} \nonumber \\ 
&&  -\frac{{\cal E}_t}{l^3}
(K_{1122}+K_{2211}+4K_{1212})\Big]
\end{eqnarray}
and where the coefficients $\Gamma_j$, $L_{ij}$, $\Lambda_{ij}$, $I_{ij}$ and 
$K_{ijkl}$ are given in appendix \ref{AppInt}.

\subsection{Modulation equations}

Applying the extended Hamilton principle (see Ref. \cite{Nayfeh04}), we obtain
the equations of motion for the modulations $A_1(T_2)$ and $A_2(T_2)$ as
\begin{equation} \label{LagEq}
\frac{d}{dT_2} \left(\frac{\partial {\cal L}}{\partial \bar{A}_i'} \right) = 
\frac{\partial {\cal L}}{\partial \bar{A}_i} + \bar{Q}_i, \ i \in \{1,2\}.
\end{equation}
From Eqs. \eqref{LagA} and \eqref{dWQ}, we then derive the following pair of 
modulation equations
\begin{multline} \label{EqsA1A2}
2i\omega_i(M_i A_i'+\mu_iA_i) = -2A_i\left(C_{ii}|A_i|^2+ \right. \\
+ \left. C_{12}|A_j|^2\right)+{\cal F}_ie^{i\sigma_i T_2},
\end{multline}
where $(i,j) \in \{1,2\}$, $i \neq j$. 
Looking for solutions in the polar form
\begin{equation} \label{PolarAi}
A_i(T_2) = \frac{1}{2} a_i(T_2)\, e^{i(\sigma_i T_2-\theta_i(T_2))},  
\ i \in \{1,2\},
\end{equation}
and separating the real and imaginary components of Eqs. \eqref{EqsA1A2},
yields
\begin{eqnarray} \label{EqsA1A2realim}
\omega_i(\sigma_i-\theta_i')M_ia_i &=& 
\frac{a_i}{4}\left(C_{ii}\,a_i^2+C_{12}\,a_j^2\right)- {\cal F}_i 
\cos(\theta_i), \nonumber \\
\omega_i(M_ia_i'+\mu_i a_i) &=& {\cal F}_i \sin(\theta_i). 
\end{eqnarray}
Looking for steady state (periodic) solutions, we impose $a_j'=0$ and $\theta_j'=0$ and we finally obtain the frequency-amplitude relations as
\begin{eqnarray} \label{FreqAmp1}
\hskip-4ex \sigma_1\! &=& \!\frac{1}{4M_1\omega_1}\left[C_{11}\,a_1^2\!+\!C_{12}\,a_2^2\right]
\pm \sqrt{\frac{{\cal F}_1^2}{M_1^2\omega_1^2a_1^2}\!-\!\frac{\mu_1^2}{M_1^2}},
\\ \label{FreqAmp2}
\hskip-4ex \sigma_2 \!&=&\! \frac{1}{4M_2\omega_2}\left[C_{22}\,a_2^2\!+\!C_{12}\,a_1^2\right]
\pm \sqrt{\frac{{\cal F}_2^2}{M_2^2\omega_2^2a_2^2}\!-\!\frac{\mu_2^2}{M_2^2}}.
\end{eqnarray} 
together with
\begin{equation} \label{Anglestheta}
\tan \theta_i = \frac{4\mu_i\omega_i}{C_{ii}\,a_i^2+C_{12}\,a_j^2-
4M_i\omega_i\sigma_i},\ i \neq j.
\end{equation} 
In first approximation the steady state solution can be cast into the form
\begin{multline} \label{yetasol}
\begin{pmatrix} 
y(x,t) \\
\eta(x,\xi,t)
\end{pmatrix}
= \varphi_1(u) \left\{ a_1 \, \begin{pmatrix} 
1 \\
\chi_1(v)
\end{pmatrix} \cos (\Omega t - \theta_1) + \right. \\ 
\left. + a_2 \, \begin{pmatrix} 
1 \\
\chi_2(v)
\end{pmatrix} \cos (\omega t - \theta_2)  \right\}.
\end{multline}
The amplitudes $a_1$ and $a_2$ are assumed small enough for the 
perturbation expansion to hold (notice that the bookkeeping parameter
$\varepsilon$ has been absorbed in the amplitudes and that Eqs. 
\eqref{FreqAmp1}-\eqref{FreqAmp2} 
can be used as such provided the detunings $\sigma_j$ 
are redefined as $\sigma_j\equiv \varepsilon^2 \sigma_j$, the viscosities
$\mu_j$ as $\mu_j \equiv \varepsilon^2 \mu_j$ and the forces ${\cal F}_j$
as ${\cal F}_j \equiv \varepsilon^3 {\cal F}_j$).

\subsection{Discussion}

The frequency-amplitude relations \eqref{FreqAmp1}-\eqref{FreqAmp2}
allow us to evaluate the frequency shift of the fundamental peak 
induced by a driving of the
higher mode at the exact linear resonance frequency, $\omega_2$. This frequency
shift is determined as the difference between the maximum
of the resonance peak of the fundamental mode and $\omega_1$. Now, the 
amplitude $a_1$ becomes maximum if the square root in the r.h.s. of 
\eqref{FreqAmp1} vanishes, that is, 
\begin{equation} \label{a1max}
a_1^{\rm max} = \left| \frac{{\cal F}_1}{\mu_1 \omega_1}\right|.
\end{equation}
For a system whose higher mode is driven exactly at frequency $\omega=\omega_2$
we have of course $\sigma_2=0$ and then the amplitude of the second peak 
is solution to 
\begin{equation} \label{a2ofa1max}
\frac{1}{4M_2\omega_2}\left[C_{22}\,a_2^2+C_{12}\,
\frac{{\cal F}^2_1}{\mu^2_1 \omega^2_1}\right]
\pm \sqrt{\frac{{\cal F}_2^2}{M_2^2\omega_2^2a_2^2}-\frac{\mu_2^2}{M_2^2}} = 0,
\end{equation}
which is a cubic equation for $a_2^2$. Once the solution 
$a_2({\cal F}_1,{\cal F}_2)$
is known, we can reinstate it in \eqref{FreqAmp1} and we obtain the 
frequency shift, $\sigma_1 = \Omega -\omega_1$ as
\begin{equation} \label{Freqshifts1}
\sigma_1({\cal F}_1,{\cal F}_2) = 
\frac{1}{4M_1\omega_1}\left[C_{11}\,\frac{{\cal F}^2_1}
{\mu^2_1 \omega^2_1}+C_{12}\,a_2^2({\cal F}_1,{\cal F}_2)\right],
\end{equation} 
which provides the frequency shift as a function of the forces 
(or driving power), ${\cal F}_1$ and ${\cal F}_2$, of the fundamental 
and excited modes.

\section{Conclusion}
Here  we have  expanded  and generalized  the  analytical modeling  of
coupled-resonator  elastic structures.  Our  one-dimensional continuum
model is shown to capture  the relevant dynamics and response spectrum
of  antenna-like  structures  with  arbitrary  (symmetric)  cantilever
distributions as well  as boundary conditions on the  central beam.  A
detailed analysis  of the band  structure of the response  spectrum in
the special cases of constant and alternating cantilever length arrays 
resonator provided insights
into the  modal dynamics of  the cantilevers versus the  carrier beam.
Design parameters such  as cantilever lengths are shown  to modify the
response spectrum and should help achieve potentially advantageous 
performance.

We further showed that perturbation  theory can be used to analyze the
nonlinear  modal  response  of  a prototype  antenna  resonator.   The
analysis was  performed without loss  of generality on two  modes that
have been selected for their practical advantages in performance.  The
anharmonic   amplitude-frequency  coupling   that  results   from  the
inclusion of nonlinear terms in the equations of motion elucidates the
experimental   behavior  of   hybrid  nano-resonators   under  strong
excitation.   Thus the  current analysis  provides a  solid analytical
foundation  to the  understanding of  device performance,  with direct
application to a wide range of current resonator devices, as well as a
forthcoming experimental report by the authors.

\begin{acknowledgments}
This work is supported by NSF (DMR-0449670).
\end{acknowledgments}

\appendix

\section{Effective coefficients} 
\label{AppInt}
We provide here the definition and the numerical/analytical evaluation of the
quantities involved in Eqs. \eqref{Param} for the antenna dimensions 
given in \cite{dorignac_JAP1}.
\begin{eqnarray} \label{integrals1} 
\Gamma_1 &=& \int_0^1 \left(\int_0^u \left[\varphi_1'(\nu)\right]^2 d\nu \right)^2 
du ,\nonumber \\ 
\Gamma_2 &=&  \int_0^1 \left[\varphi_1'(u)\varphi_1''(u)\right]^2 du,
\end{eqnarray}

\begin{eqnarray} \label{integrals2} 
\Gamma_3 &=&  \int_0^1 \varphi_1^4(u)\, du,\nonumber \\
\Gamma_4 &=&  \int_0^1 \varphi_1(u)\, du,\nonumber \\
L_{ij} & =& \int_0^1 h_i(v)h_j(v) \, dv,\nonumber \\  
\Lambda_{ij} & =& \int_0^1 h_i(v)(h_j(v)-1) \, dv,\nonumber \\ 
I_{ij} &=& \int_0^1 \left(\int_0^v h_i'(\nu) h_j'(\nu) \, d\nu \right)^2 dv, 
\nonumber \\  
K_{ijkl} &=&\int_0^1 h_i'(v) h_j'(v) h_k''(v) h_l''(v) \,dv.  
\end{eqnarray}
with $h_j(v) \equiv \chi_j(v)+1$. 
All the integrals above can normally be evaluated analytically. 
For $\Gamma_i$, $i \in \{1,\dots,4\}$, results are simple
enough to be displayed below. 
\begin{eqnarray}
\Gamma_1 & = & \frac{\beta_1 t}{2}(\beta_1 t+2) \simeq 6.1513, \nonumber \\
\Gamma_2 & =&  \frac{\beta_1^5t}{10}(5\beta_1t+11) \simeq 2846.4975, \nonumber \\
\Gamma_3 & =&  \frac{3}{4}\left(3-t^4-\frac{2t^3}{\beta_1}\right) 
\simeq 1.8519, \nonumber \\ 
\Gamma_4 &=& \frac{4t}{\beta_1} \simeq -0.8308 \nonumber 
\end{eqnarray}
where $t =\tan (\beta_1/2)$. Note that $\cos (\beta_1)=1/\cosh(\beta_1)$
and $\sin(\beta_1)=-\tanh(\beta_1)$.
Apart from $\Lambda_{ii}=L_{ii}-2A_2(\gamma_i)/\gamma_i$ and
$L_{ii}\equiv L(\gamma_{1,i})$, 
the other integrals have been 
evaluated numerically for the parameters given in \cite{dorignac_JAP1}. 
We have found
\begin{eqnarray}
& &  L_{11} \simeq 1.00012 \ ;\ L_{22} \simeq 3.89887, \nonumber \\
& &  \Lambda_{11} \simeq 6.16\times 10^{-5} \ ;\ \Lambda_{22} \simeq 4.9687, 
\nonumber \\
& & I_{11} \simeq 1.64 \times 10^{-16} \ ;\ I_{22} \simeq 232.49, \nonumber \\
& & I_{12} \simeq 1.94 \times 10^{-7}\ ;\ K_{1212} \simeq 8.26 \times 10^{-7}, 
\nonumber \\
& &  K_{1111} \simeq 6.58 \times 10^{-16}\ ;\
K_{2222} \simeq 1.07 \times 10^{3}, \nonumber \\
& & K_{1122} \simeq 9.84 \times 10^{-7} \ ;\  
K_{2211} \simeq 6.98 \times 10^{-7}. \nonumber 
\end{eqnarray}
So that, finally,
\begin{eqnarray} \label{Paramnum}
&& M_1 = 1.74 \times 10^{-14}\ \ ;\ \  M_2 = 4.17 \times 10^{-14}, \nonumber \\
&& {\cal F}_i = -4.44 \times 10^{-6} f_i \ \ ;\ \ 
C_{12} = 1.65 \times 10^{17}, \nonumber \\
&& C_{11} = -5.07 \times 10^{13} \ \ ;\ \ C_{22} = 1.05 \times 10^{21}, 
\nonumber \\
&& \omega_1 = 1.55 \times 10^{8}\ \ ;\ \  \omega_2 = 1.85 \times 10^{10}. 
\nonumber 
\end{eqnarray}
where all the results are given in SI units.
From the quantities $\omega_j$, the frequencies of the 
fundamental and collective modes are $24.7$ MHz and $2.94$ GHz, 
respectively.

\end{document}